# Atomically sharp interface enabled ultrahigh-speed, nonvolatile memory devices

Liangmei Wu[1,2,7], Aiwei Wang[1,2,7], Jinan Shi[2,7], Jiahao Yan[1,2,7], Zhang Zhou[1,2], Ce Bian[1,2], Jiajun Ma[1,2], Ruisong Ma[1,2], Hongtao Liu[1,2], Jiancui Chen[1,2], Yuan Huang[1], Wu Zhou[2], Lihong Bao[1,2,3]*, Min Ouyang[4]*, Stephen J. Pennycook[5], Sokrates T. Pantelides[6,2], Hong-Jun Gao[1,2,3]*

[1] Institute of Physics, Chinese Academy of Sciences, P. O. Box 603, Beijing 100190, PR China

[2] University of Chinese Academy of Sciences & CAS Center for Excellence in Topological Quantum Computation, Chinese Academy of Sciences, P. O. Box 603, Beijing 100190, PR China

[3] Songshan Lake Materials Laboratory, Dongguan, Guangdong 523808, PR China

[4] Department of Physics, University of Maryland, College Park, MD 20742, US

[5] Department of Materials Science and Engineering & Centre for Advanced 2D Materials, National University of Singapore, 117575, Singapore

[6] Department of Physics and Astronomy & Department of Electrical Engineering and Computer Science, Vanderbilt University, Nashville, Tennessee 37235, US

[7] These authors contributed equally: Liangmei Wu, Aiwei Wang, Jinan Shi, Jiahao Yan.

*Correspondence to: hjgao@iphy.ac.cn, lhbao@iphy.ac.cn, mouyang@umd.edu.

**Development of memory devices with ultimate performance has played a key role in innovation of modern electronics. As a mainstream technology nonvolatile memory devices have manifested high capacity and mechanical reliability, however current major bottlenecks include low extinction ratio and slow operational speed. Although substantial effort has been employed to improve their performance, a typical hundreds of micro- or even milli- second write time remains a few orders of magnitude longer than their volatile counterparts. We have demonstrated nonvolatile, floating-gate memory devices based on van der Waals heterostructures with atomically sharp interfaces between different functional elements, and achieved ultrahigh-speed programming/erasing operations verging on an ultimate theoretical limit of nanoseconds with extinction ratio up to $10^{10}$. This extraordinary performance has allowed new device capabilities such as multi-bit storage, thus opening up unforeseen applications in the realm of modern nanoelectronics and offering future fabrication guidelines for device scale-up.**

Memory technology has played a pivotal role in the development of semiconductor industry in the past decades, driven by the explosive growth of massive data storage and desire of ultrafast data processing[1]. Current bottleneck in memory field includes operation speed, data retention, endurance and extinction ratio[1,2]. In particular, while the scaling of devices continues, silicon-based technology will soon reach a critical limit to meet the increasing demands for memory capacity. One of the key challenges is related to the unavoidable interfacial dangling bonds in ultrathin-body silicon, which limit electrostatic control of the channel and lead to high power consumption caused by increasing off-state current[3]. It is thus an urgent need to seek defect-free interfaces and integrate them into the device architecture. Among all candidates, the emerging two-dimensional (2D) materials[4-8] and their heterostructures[9-13] are potentially free from surface dangling bonds and immune to short-channel effects that can allow effective electrostatic control and mechanical flexibility[14]. Indeed, a few examples employing 2D materials for flash memory devices have been recently attempted[15-26] but with limited device performance. For example, a very long write time on order of milliseconds to seconds was observed in the 2D materials based floating-gate memory devices[15-18], while an alternative semi- floating-gate configuration has shown improved write time but extremely short retention time of ~10 s[27], making it unsuitable for long-term storage. Theoretically, an ideal floating-gate memory devices based on planar layer materials should allow nanosecond order operational time (Supplementary Figs. 1 and 2), but the key is control of interfacial quality in a practical device that has not yet been addressed so far.

Here, we demonstrate that nonvolatile floating-gate memory devices with exceptional performance verging on an ultimate theoretical limit can be achieved without requirement to modify commercial device architecture by employing 2D van der Waals heterostructures with improved interfacial coupling. In particular, ultrahigh-speed operation with nanosecond write and read times that is limited by instrumentation response, extremely high extinction ratio of $10^{10}$, and a 10-yr retention time have been successfully demonstrated. Our result represents a leap forward to understand current bottleneck of memory devices as well as future memory and storage technologies.

Figure 1a (top) shows a schematic of our floating-gate field-effect transistor (FGFET) device, consisting of a vertically stacked InSe/hBN/multilayer-graphene (MLG) van der Waals heterostructure on a $SiO_2/p^{++}$ Si substrate, where InSe, hBN, MLG, $SiO_2$, and $p^{++}$ Si, serve as channel, tunnel-barrier, floating gate, control-gate dielectric, and control gate, respectively. Figure 1b shows a false-color optical image of a typical device in which one 2.5-nm-thick MLG (marked by blue dashed line) is placed directly on 300 nm $SiO_2$/ Si substrate, followed by sequential stacking of 15.8-nm-thick hBN (marked by white dashed line) and 14.4-nm-thick InSe on top of the MLG. Source and drain electrodes are fabricated using standard e-beam lithography followed by thermal evaporation. The detailed materials preparation and device fabrication processes are provided in Methods and Supplementary Fig. 3. The MLG floating gate can be charged/discharged by an electric field, and charges can be retained in the MLG after removing the electric field (bottom, Fig. 1a), regulating the conductivity of the channel materials to fulfil nonvolatile memory function[28]. Our device fabrication is general and reproducible. For example, devices with same configuration and structural quality have been also achieved by using other 2D layers as channel material, including $MoS_2$.

We have employed an aberration-corrected scanning transmission electron microscope (STEM) to thoroughly characterize interfaces between different functional components in as-fabricated devices. Figure 1c and 1d present typical low- and high- resolution atomic-number-contrast (Z-contrast) high-angle-annular-dark-field (HAADF) STEM cross-sectional images of one device, respectively. More bright-field and HAADF STEM images are presented in Supplementary Fig. 4. Our results have unambiguously revealed the existence of extremely uniform and clean atomically sharp interfaces between different functional layers in our devices, without any gap or observable defects/contamination (Figs. 1c and 1d). These features are the key for the exceptional device performance presented below.

Figure 2a shows typical transfer curves of an InSe FGFET at room temperature. The data were acquired by sweeping control-gate voltage ($V_{cg}$) from negative to positive values and then back to negative values, with the MLG floating and a fixed drain-source bias ($V_{ds}$) of 0.05 V. A sizeable hysteresis is clearly observed in Fig. 2a. We have performed two control experiments by measuring the same FGFET devices with grounded MLG (Supplementary Fig. 5a) and by fabricating independent devices without the MLG layer (Supplementary

Fig. 5b). Both controls have shown negligible hysteresis in their transfer curves, indicating a low trapped-charge density at the InSe/hBN interface that is consistent with our structural characterization of exceptional interface quality in the InSe/hBN heterostructure. By comparing with result in Fig. 2a, we conclude that the presence of the large voltage hysteresis can be attributed to the existence of the MLG floating gate. The appearance of substantial hysteresis signifies the storage capability of a memory device, and is characterized by a memory window width ($\Delta V$) that is defined as the shift of the threshold voltage in the dual-swept $V_{cg}$. In general, a larger memory window suggests that more charges can be stored in the MLG floating gate, resulting in a more reliable memory-device performance (for example, a minimum width of 1.5 V is typically necessary in order to produce a reasonable on/off ratio for reliable memory application[19]). Figure 2b shows that the $\Delta V$ increases proportionally with the maximum value of control-gate voltage ($V_{cg,max}$), whereby the amount of charge stored in the MLG can be tuned by the applied $V_{cg}$. The complete set of data is shown in Supplementary Fig. 6. The $\Delta V$ can be as high as 64 V when the $V_{cg}$ is swept between -40 V and +40 V, which is higher than all previous reported memory devices based on 2D materials[15-19] (Supplementary Table 1). The corresponding charge density stored in the MLG floating gate can thus be estimated by $(\Delta V \times C_{CG\text{-}FG})/q$, where $q$ is the electron charge of $1.6 \times 10^{-19}$ C and $C_{CG\text{-}FG}$ is the dielectric capacitance between the control gate and the floating gate (300 nm $SiO_2$) of $1.15 \times 10^{-8}$ F·cm$^{-2}$, resulting in a value of $4.6 \times 10^{12}$ cm$^{-2}$. Such large charge density achievable in the MLG floating gate

outperforms polycrystalline silicon in conventional flash memory, and can facilitate easy detection in practical applications[28]. There are a few advantages of utilizing MLG as the floating gate, including a significant reduction of floating-gate interference[29,30] and effective suppression of ballistic currents across the floating gate due to the low conductivity along the c-axis. Furthermore, as compared with single-layer graphene, MLG has a higher work function (4.6 eV) that is independent of the number of layers, enabling longer charge retention time as well as higher electronic density of states[19]. All these attributes of MLG have contributed to the observed large memory window in our devices.

We have proposed and defined in Fig. 3a the programming and erasing processes achievable in our FGFET devices for a nonvolatile floating-gate memory: When a positive voltage pulse is applied to the control gate with the drain and source grounded and the MLG floated (step *i* in Fig. 3a), electrons from the InSe channel tunnel through the hBN barrier and accumulate in the MLG, completing a program operation. The accumulated electrons in the MLG are well retained even after removing the external electric field due to low return probability, resulting in a positive shift of the threshold voltage and thus low conductance of the InSe channel at $V_{cg} = 0$ V. This state is defined as "Program State" (Step *ii* in Fig. 3a). On the contrary, when applying a negative voltage pulse to the control gate (step *iii* in Fig. 3a), stored electrons can tunnel from the MLG floating gate back to the InSe channel, i.e. erase operation, leading to a negative shift of the threshold voltage after the external electric field is turned off. As a result, the InSe channel manifests high conductance at $V_{cg} = 0$ V, which is defined as "Erase State" (Step *iv* in Fig. 3a).

Figure 3b shows a typical realization of the proposed memory operations in our devices by using +17.7/-17.7 V voltage pulses with 160 ns full width at half maximum (FWHM) for programming/erasing operations (the output voltage pulse waveforms are shown in Supplementary Fig. 7). When a positive square pulse is applied to the control gate, the InSe channel is driven to a low-conductance state, i.e. program state. This program state can be "erased" by applying a negative voltage pulse to the control gate to recover the high conductance of the InSe channel, i.e. erase state. It is worth noting that the amplitude of the programming and erasing voltage pulses in Fig. 3b (17.7 V) is comparable to that of commercial flash memory (~15 V)[1], whereas the extinction ratio between the erase and program states is up to $10^{10}$ that is much higher than other floating-gate memory devices[15-18] (Supplementary Table 1). This high extinction ratio not only allows facile and reliable readout of the memory states to reduce data sensing error, but also opens up other attractive applications such as multi-bit storage in the same memory cell, which is challenging otherwise.

Our InSe floating-gate memory devices are very robust with high endurance and high reliability. We have extracted the threshold voltage of the InSe channel from transfer curves at different time intervals (Supplementary Fig. 8), and summarized the result in Fig. 3c. This result suggests an 80.2% difference remaining in threshold voltage between the program state (red dashed line) and the erase state (blue dashed line) even after 10 years, thus making long-term data retention become feasible at room temperature. The endurance test by repeatedly programming and erasing the same memory cell has also been performed with result presented in Fig. 3d. It can be clearly seen that both program and erase states remain almost unchanged with negligible degradation even after 2000 cycles. The robust retention and endurance performance of our memory devices can be attributed to the defect-free hBN tunneling barrier layer integrated in our devices[18,31].

Importantly, we have demonstrated that our InSe floating-gate memory devices feature ultrahigh-speed operation, which is not available in conventional flash-memory devices. Figure 4a shows that the InSe floating-gate memory device can be programmed/erased successfully by an ultrashort +20.2/-20.8 V voltage pulse down to 21 ns FWHM. The transient responses of memory devices by following the ultrashort programming and erasing pulses are shown in Figs. 4b and 4c, respectively. A few important characteristics can be immediately identified under ultrafast operations: (1) The data storage and erasure can still be consistently achieved with high extinction ratio up to $10^{10}$ even with ultrashort programming and erasing pulses; (2) The memory states can respond instantaneously and settle without delay. We have defined the response time of a device as the time required for it to reach 1/e value of its desired state after the peak pulse, and estimated a 36 ns and 43 ns response time for programming and erasing operations with a FWHM pulse width of 21 ns, respectively. It is worth noting that these response time estimations should be taken as the upper bound values and are currently limited by voltage pulse source available in our work. Nevertheless, the observed ultrahigh-speed operation of floating-gate memory devices has approached theoretical prediction of an ideal device possessing the same experimental device parameters (Supplementary Fig. 2), and is ~5000 times faster than that of commercially produced flash-memory devices and comparable with that of commercially available volatile DRAM products. This is again consistent with our confirmation of atomic sharp interfacial feature in Figs. 1c and 1d. Because of the extremely fast transient response for both programming and erasing operations, it can allow a true ultrahigh speed memory device operation. Figure 4d shows one example of ultrahigh frequency operation of memory cell by sequential programming and erasing up to 5 MHz that is again limited by instrumentation response; and (3) our fabrication of layer materials based floating-gate memory devices with atomic sharp interfaces can be reproducibly extended to other 2D materials, with one example shown in Supplementary Fig. 9 by utilizing $MoS_2$ as channel, highlighting the generality and reproducibility of the roles of atomically sharp interfaces in 2D memory technology.

The combination of high extinction ratio and ultrafast operation characteristics achievable in our van der Waals memory devices immediately open up opportunities for novel device configurations and applications. For example, the current extinction ratio of $10^{10}$ should allow achievement of multi-bit memory operation to store more than one bit of information per cell for ultrahigh density information storage. Nevertheless, in a commercial multi-level cell flash memory, multi-bit memory operation is often achieved by controlling the amplitude of $V_{cg}$ to change the amount of charge stored in the floating gate, requiring extra control of memory cells[32]. Enabled by ultrafast operation capability demonstrated in Fig. 4a, an alternative mechanism of multi-bit storage can be achieved by a judicious design of the pulse sequence to control the amount of charges stored in the MLG floating gate, offering more flexibility for data programming. Figure 4e exemplifies realization of a two-bit storage in our InSe-based memory cell. In this demonstration we have utilized same ultrashort pulses with FWHM of 21 ns in Figs. 4b and 4c for programming and erasing. More specifically, by applying a pre-defined pulse sequence we have shown that well discernable multiple states of memory cells can be achieved reproducibly. For example, Figure 4e shows that the memory cell can be programmed to different program states of (10), (01), (00) with sufficient large extinction ratio and good retention performance after two, three, and four positive voltage pulses (+20.2 V, 21 ns), respectively. It should be noted that in this demonstration for every programming operation an erase operation is applied to reset the memory state to (11), which has been consistently achieved by applying a single negative voltage pulse (-20.8 V, 21 ns). Furthermore, by optimizing our memory devices (for example, the thickness of the hBN tunnel-barrier layer), higher storage capacity such as Triple Level Cell and Quadruple Level Cell has also been demonstrated with reliable performance (Supplementary Figs. 10 and 11). Such multi-bit storage scheme can only be possible when both high extinction ratio and ultrafast operation become available, highlighting the uniqueness of our demonstrated memory device configuration and performance.

We have fabricated and exemplified emerging nonvolatile floating-gate memory devices based on 2D functional layer materials, exhibiting device performance approaching the theoretical limit of an ideal memory structure and exciting ultrahigh speed operation for the first time, and the key is our achieved atomically clean and sharp interfaces inherent in the as-fabricated memory devices. Our findings and device performances have important implications and opens up exciting opportunities on several fronts. Our device configuration follows the commercial floating gate memory architecture but with a few orders of improvement in the device performance. As a result, it should facilitate adoption of existing industry standards with emerging technology. Although current device fabrication is based on exfoliation method, our finding of roles of atomically sharp interface between functional layers in the performance limit of memory devices should offer a fabrication guideline for future integration with other techniques for the device scale-up[33-37]. Moreover, the success of ultrahigh-speed nanosecond operation and response in our as-fabricated devices has solved the long-term challenge in the field of nonvolatile memory technology, and is comparable to commercial volatile DRAM technology, representing a big leap forward in the current demanding data storage and data processing. In addition, such outstanding performance should allow many new unforeseen operational schemes, including multi-bit ultrahigh-density information storage demonstrated in the present work, which should be crucial for recent development of big data science. All these together offer the ground of next-generation electronic devices based on emerging 2D materials with optimum device engineering.

## Methods

**Device fabrication and electrical characterization.** InSe, $MoS_2$, hBN, and graphite bulk crystals are purchased from 2D Semiconductors, SPI Supplies, HQ graphene, and NGS Naturgraphit companies, respectively. InSe/hBN/MLG and $MoS_2$/hBN/MLG heterostructures were prepared on a silicon wafer with 300 nm $SiO_2$ using mechanical exfoliation and the dry-transfer approach[38,39], carried out in an argon atmosphere glovebox with the concentrations of $O_2$ and $H_2O$ below 0.5 ppm. Then Cr/Au drain and source electrodes were defined by standard e-beam lithography, thermal evaporation, and lift-off. The floating-gate memory devices were protected by a layer of spin-coated PMMA at once after lift-off[40]. Electrical measurements were performed with a Keithley 4200 semiconductor characterization system (4200-SCS) and a home-made electric circuit in a vacuum probe station at room temperature. Memory characterization of +17.7/-17.7 V voltage pulses with 160 ns FWHM were performed in 4200-SCS equipped with 4225-PMU and 4225-RPM. The ultrahigh-speed characterization of InSe and $MoS_2$ floating-gate memory devices was executed with the control-gate terminal connected to a home-made nanosecond voltage pulse signal while the drain and source terminals were connected to 4200-SCS source-measure units. The grounds of the home-made electric circuit and the 4200-SCS needed to be connected together. The response time tests were executed in atmospheric environment. The drain-source current is amplified with a FEMTO DHPCA-100 (gain of $10^4$ V/A and band width of 14 MHz). The nanosecond voltage pulse and amplified drain-source current were recorded with a Tektronix MDO4104B-3 Mixed Domain Oscilloscope (band width of 1GHz). The thicknesses of InSe, hBN and MLG flakes were determined by a Bruker Dimension Edge atomic force microscope after all tests.

**STEM sample preparation and characterization**. The InSe/hBN/MLG heterostructure on a $SiO_2$/Si substrate was protected by covering with three layers of graphite flakes with total thickness reaching dozens of nanometers by the dry-transfer approach before TEM sample fabrication. Then a standard cross-section TEM lift-out method was carried out with an FEI Helios NanoLab G3 CX focused-ion-beam microscope for high-resolution imaging. The HAADF and bright-field (BF) imaging were performed in an aberration-corrected Nion HERMES-100 STEM at 60 kV. Atomic-number-contrast images were obtained using an illumination semi-angle of 32 mrad, and collection semi-angle from 75 to 210 mrad for HAADF imaging and within 9 mrad for bright-field imaging.

## Acknowledgments

We thank Yu-Yang Zhang, Shixuan Du, Guojian Qian, and Zhili Zhu for helpful discussions, and Haifang Yang, Junjie Li, Changzi Gu, and Qing Huan for assistance in device fabrication and measurement. This work was supported by National Key Research & Development Projects of China (Grant No. 2016YFA0202300 & 2018FYA0305800), National Natural Science Foundation of China (Grant Nos. 61674170, 61888102), Strategic Priority Research Program of Chinese Academy of Sciences (CAS, Grant No. XDB30000000, XDB28000000), Youth Innovation Promotion Association of CAS (20150005), and the CAS Pioneer Hundred Talents Program. M.O.Y. acknowledges support from ONR (N000141712885) and NSF (DMR1608720). S.J.P. acknowledges support from the Ministry of Education, Singapore, under a Tier 2 grant (No. MOE2017-T2-2-139). A portion of the research was performed in the CAS Key Laboratory of Vacuum Physics.

## Author contributions

H.-J.G. supervised the overall research. L.H.B., M.O.Y, and H.-J.G. designed the experiments. L.M.W., A.W.W., J.H.Y., and L.H.B. fabricated the devices and carried out electrical measurements. J.A.S., S.J.P., and W.Z. performed the STEM measurements. A.W.W. constructed the home-made electric circuit with ultrashort-voltage-pulse signals with 21 ns FWHM. L.M.W., L.H.B., A.W.W., J.A.S., J.H.Y., W.Z., M.O.Y., S.T.P., and H.-J.G. analyzed the data. L.M.W., L.H.B., J.A.S., W.Z., S.J.P., S.T.P., M.O.Y., and H.-J.G. wrote the paper. All authors contributed to the preparation of the manuscript.

## Competing interests

The authors declare that they have no competing interests.

# Figure legends

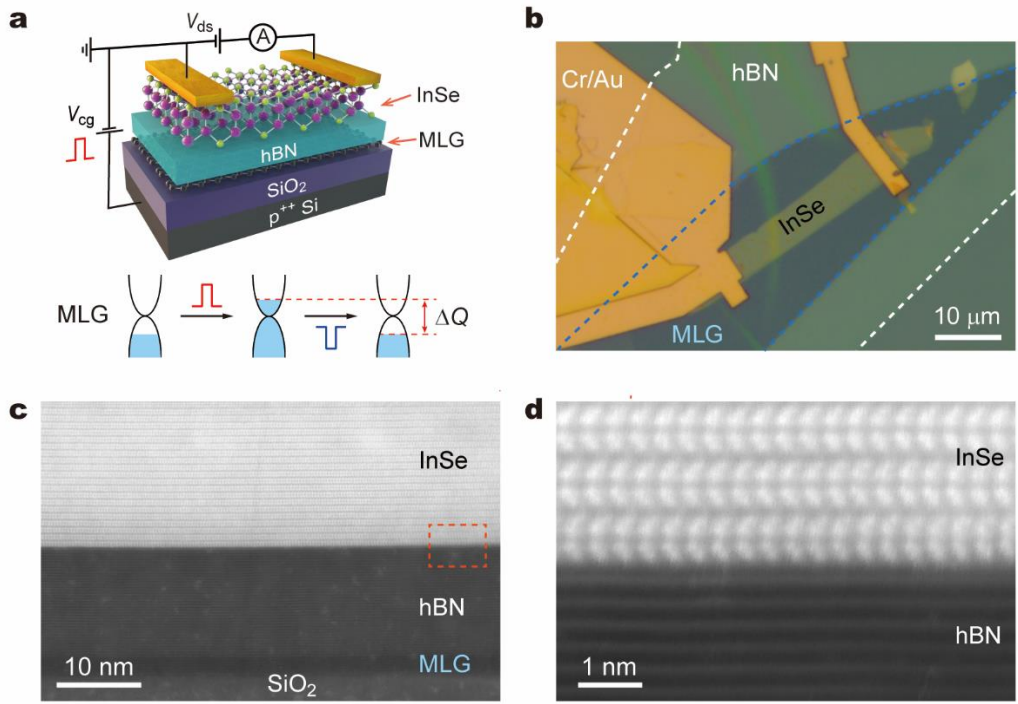

**Fig. 1 | Atomically sharp interfaces embedded in nonvolatile memory device with floating-gate configuration built upon van der Waals heterostructures. a,** Schematic of a floating-gate memory device configuration and its operational principle. **b,** False-color optical image of a memory device based on InSe/hBN/MLG heterostructure placed on a SiO$_2$/Si substrate. **c,d,** low- (**c**) and high- (**d**) resolution HAADF-STEM images of a InSe/hBN/MLG heterostructure on SiO$_2$, highlighting uniform, clean, and atomically sharp 2D interfaces without defects and contamination between InSe, hBN and MLG, respectively. The image in (**d**) corresponds to the orange rectangular dotted area in (**c**).

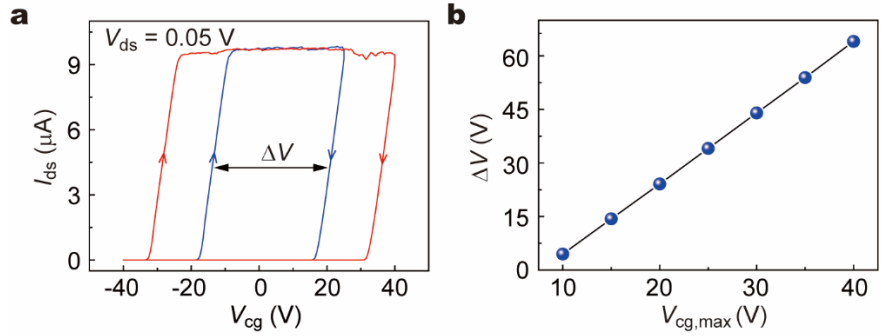

**Fig. 2 | 2D InSe based floating-gate memory device with large memory window. a,** Typical dual-sweeping transfer curves with FGFET configuration. The data were acquired by sweeping the $V_{cg}$ with the MLG floating, while monitoring the $I_{ds}$ under fixed $V_{ds}= 0.05$ V. The blue and red curves correspond to the maximum $V_{cg}$ of 25 V and 40 V, respectively. Both transfer curves clearly reveal the existence of large $\Delta V$. The arrows show sweep directions of the $V_{cg}$. **b,** Linear dependence of $\Delta V$ on $V_{cg,max}$.

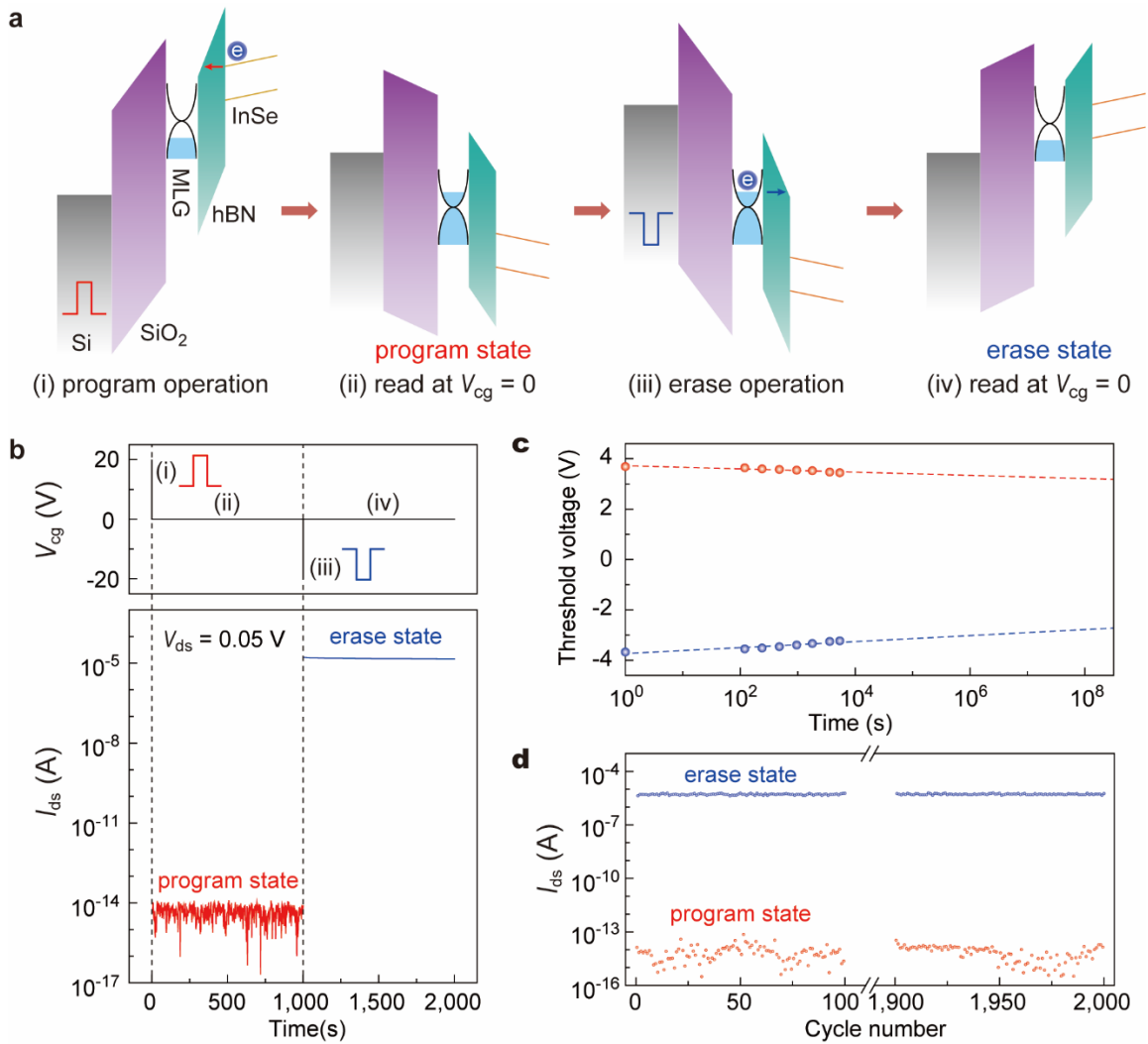

**Fig. 3 | Programming and erasing nonvolatile InSe floating-gate memory devices with large extinction ratio and robust performance. a,** Schematic energy-band diagrams of a memory device for programming, reading and erasing operations. **b,** The conductance of the InSe channel can be toggled between low-conductance state (i.e. program state) and high-conductance state (i.e. erase state) by applying a single +17.7 V or -17.7 V voltage pulse with 160 ns FWHM, respectively. **c,** Time variation of threshold voltage after +17.7/-17.7 V programming/erasing pulses with 160 ns FWHM. **d,** Endurance performance test of the memory device executed with alternative voltage pulses (+17.7/-17.7 V programming/erasing pulses with 160 ns FWHM).

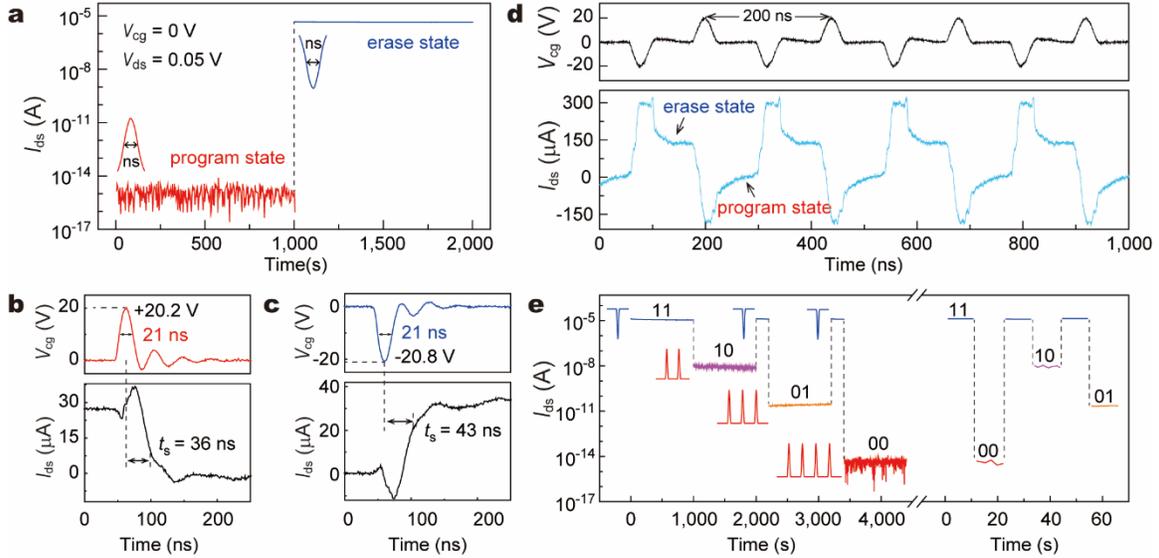

**Fig. 4 | Ultrafast operation of memory cell and enabled multi-bit storage paradigm. a,** The InSe memory cell can be successfully programmed/erased by a positive/negative voltage pulse with nanosecond pulse width, while maintaining high extinction ratio of $10^{10}$. The amplitude of voltage pulse is +20.2 V/-20.8 V. **b,c,** Ultrafast response of memory cells after programming (**b**) and erasing (**c**) pulses. **d,** Demonstration of ultrahigh frequency operation by sequential programming and erasing pulses with ~100 ns intervals. **e,** Realization of reproducible multi-bit storage by a simple combination of ultrafast pulse sequence, with an example of repeatable 2-bit memory capacity demonstrated in the figure. By designing and applying sequential pulses for data programming, multiple level states of the memory cell can be programmed with discernible extinction ratio to achieve multi-bit storage. All programmed states can be consistently erased by a single nanosecond negative voltage pulse. The voltage amplitude and FWHM of programming and erasing pulses are (+20.2 V, 21 ns) and (-20.8 V, 21 ns), respectively.